# *Arknights:* Playable Explanation and Player Agency under Opacity

**Shuai Guo**
Department of Informatics and Media
Uppsala University, Uppsala, Sweden
shuai.guo.5671@student.uu.se

**ABSTRACT**
As generative AI increasingly mediates learning and decision-making, users often act effectively while struggling to interpret how system outcomes are produced. While Explainable Artificial Intelligence (XAI) research has primarily addressed this problem through transparency and visualization, less attention has been paid to how explanation is constructed through interaction. This paper examines digital games as explainable interfaces by analyzing how explanation can be configured as a playable process. Using *Arknights* as a case study, the paper conducts a qualitative close reading and interface analysis of the diegetic AI system PRTS, focusing on the implied player. The analysis shows that PRTS provides "usable but unverifiable" explanations: sufficient to initiate action, yet insufficient to stabilize causal understanding. Through incomplete information, delayed feedback, and narrative disruptions of trust, player agency is reorganized from direct control toward interpretive and abductive reasoning. The paper conceptualizes this mode as explanatory agency and discusses its implications for XAI-oriented interface design.

**INTRODUCTION**
With the continuous emergence of easy-to-use artificial intelligence technologies, students at various educational stages are increasingly able to rely on such technologies to assist their learning. However, a new learning crisis has gradually been observed within global education systems: while AI assistance enables students to efficiently complete learning tasks, they often find it difficult to fully trace their own thinking process and to explain the reasoning behind their answers. As one American high school student frankly pointed out in an article in *The Atlantic*: "When you can hand thinking over to a machine and still get an A, who cares whether you actually learned anything?" (Rosario, 2025). AI has created an efficient yet blind cognitive environment: we can quickly reach the destination, yet struggle to understand how the path toward it is generated, and may even fail to realize that we "have not truly understood." Therefore, the requirement of transparency in generative artificial intelligence technologies has been proposed, suggesting that all stages of AI decision-making processes should be open, inspectable, and explainable (Felzmann et al., 2019). However, in reality, the opacity of many AI systems does not necessarily result from intentional concealment by designers, but instead arises from the limitations of the technology itself—particularly those based on deep neural network models.

At present, Explainable Artificial Intelligence (XAI) seeks to open the "AI black box" to reveal how AI operates, thereby cultivating learners' interpretive reasoning abilities when interacting with complex machine learning systems. Unlike educational systems that attempt to improve transparency through visualization or rule exposure, digital games often embed understanding within consequential actions. Players do not passively receive explanations; rather, they make decisions under conditions of uncertainty, gradually forming an understanding of the system through failure, adjustment, and repeated experimentation. Especially in strategy-oriented games, understanding is no longer a prerequisite, but a process gradually constructed through gameplay. This makes digital games an ideal medium for examining how explanation, understanding, and agency interweave within opaque systems.

This paper takes the PRTS system in Arknights as an example to explore a design approach that presents AI explanatory structures through the configuration of player agency. It analyzes how PRTS, as an embodied diegetic AI representation, dynamically shapes players' decision-making and exploratory agency in complex systems by both providing and blocking explanation.

Before proceeding to the analysis, it is necessary to clarify several points in order to avoid any misunderstanding about the scope and intention of this paper. This paper does not claim that a game such as Arknights has any direct conceptual connection with XAI functional models, nor does it suggest that the latter directly inspired the game developers. Rather, I draw on the conceptual toolkit provided by XAI and player agency research. However, this does not exclude the possibility that other cultural theoretical frameworks might offer alternative or even better tools to address the research questions raised above. In addition, it should be emphasized that the "black box" discussed in this paper specifically refers to a phenomenological black box constructed through interface, narrative, and feedback, rather than algorithmic logic in the sense of software engineering. This distinction ensures that the focus remains on "explanatory behaviors within human-AI interaction," rather than on the technical implementation of AI.

This paper focuses on how opaque AI representations in games are designed as playable objects of interpretation, and how players, in interacting with these representations, gradually develop an understanding of the system through abductive reasoning. Although existing game studies have discussed agency, opacity, and rule comprehension, there is still a lack of systematic analysis of how "interface-based AI representations" mediate players' explanatory agency. Therefore, this paper is guided by the following two research questions:

**RQ1:** In Arknights, how does PRTS, as a diegetic AI representation, mediate players' modes of action and configure their player agency?

**RQ2:** Under conditions of system opacity, how do the interpretive gaps generated by PRTS influence players' attribution of action outcomes, their calibration of trust toward the system, and their overall path toward understanding?

## LITERATURE REVIEW

### From "Freedom" to Relational Agency

"Player Agency" is a core concept widely used in game studies and game design, and is commonly traced back to Janet Murray's theoretical formulation. In her seminal work, Murray provides an influential definition of player agency as "the satisfying power to take meaningful action and see the results of our decisions and choices" (Murray, 1997).This definition distinguishes player agency from mere participation or operation, emphasizing the relationship between intention and systemic consequences. However, this definition has also been revised and critiqued by subsequent research. Wardrip-Fruin et al. (2009) argue that Murray's formulation has often led agency to be misunderstood in practice as a subjective experience grounded in free will, while overlooking the role of game systems in shaping players' intentions and possible actions. Agency should instead be understood as a relational phenomenon involving both the game system and the player, emerging when an alignment is formed between the player's intended actions and the set of actions made available by the system. In other words, agency is not about "how much freedom the player has," but about "how the system organizes the space of action and enables players to establish correspondences between intention and outcome within it."

From this perspective, the production and experience of agency are organized within the magic circle constructed by game designers (Huizinga, 1980). When we enter the magic circle, we step into a fictional game world—a temporary space governed by its own distinctive rules—even though our bodies remain in the real world. Within this space, we often feel that we possess greater interactivity and choice than in everyday life, and experience a sense of agency as "having an impact on the game world."

### *Action without Understanding: The Limits of Action-Centered Agency*

The above account seems to imply a premise: that before entering the magic circle, players have already reached some form of tacit agreement with the game system and possess basic knowledge about the game world, enabling them to act smoothly between rules and feedback. Yet empirical experience suggests otherwise. Every player must have encountered their very first game at some point in life; even highly experienced players continue to encounter mechanisms and situations they have never seen before. In such moments, players' understanding of the game world has not yet taken shape, and the meaning of their actions often emerges only gradually through repeated trial and error. If "smoothly taking meaningful action" is treated as a prerequisite for agency, does this mean that players' agency is somehow "invalid" or "out of control" in these moments? This question suggests that agency may not be grounded in prior understanding of the system, but is instead gradually shaped and experienced through sustained interaction with non-understanding, uncertainty, and even confusion.

 C. Thi Nguyen (2019) offers a more radical interpretation of player agency. Games are not merely sites where agency occurs; rather, they are a cultural form that uses agency as its primary expressive material. Through the configuration of goals, abilities, and constraints, game designers "sculpt" an agential skeleton within which players temporarily inhabit particular modes of action. Players do not simply execute operations in games; by voluntarily taking on the game's goal structure, they treat goals that would otherwise be insignificant as provisional ultimate ends, and in doing so enter a form of layered agency. At the outer layer, the self can always exit the game, but at the inner layer, the "in-game self" must pursue victory or goal completion with seriousness and commitment for agency to be fully experienced. However, it is important to note that this action-centered account of agency still assumes that players are able to establish intelligible causal links between actions and outcomes. When facing highly opaque systems or systems in which information is heavily mediated and filtered, this assumption does not always hold: players may still be able to act, yet struggle to form a stable explanatory framework. As a result, the experience of agency shifts from "how to act" to "how to understand the system within which one is acting." From this perspective, constraints, incomplete information, and the risk of failure do not necessarily weaken agency; instead, they may serve as materials for "sculpting" action experience. They compel players to develop strategies under limited means, make trade-offs, and interpret success or failure as consequences of their own judgments, thereby strengthening the perceptibility and attributability of agency.

This also suggests that player agency should not be simply equated with interactivity. Almost all interactive systems allow users to input commands and receive feedback, yet interactivity alone is insufficient to constitute agency. Unlike functional or efficiency-oriented software, many games do not prioritize optimal paths or task completion. Instead, they often place players within cycles of "action–observation–adjustment" through rule constraints, incomplete information, and delayed feedback, enabling players to form meaningful decisions under conditions of uncertainty and trial and error. Therefore, the key to player agency lies not in "whether there is an input channel," but in how games organize the predictability of action, allowing players to understand their choices as meaningful actions and to confirm or revise this understanding through system feedback.

### *Opacity, Interpretation, and Explanatory Agency*

Player agency is also closely related to the opacity of game systems and the difficulty of interpreting them. Vella (2015) argues that there is always a gap between the objective game system—as an opaque, irreducible, and user-independent structure—and the game as an object of player interpretation. Players must make conceptual efforts to understand rules, processes, and mechanics, and this separation itself constitutes the core experience of the "ludic sublime": a tension between phenomenal experience and ontological unknowability. Even though a game's algorithms and internal properties hidden in code remain inaccessible to players, we can still analyze how game objects at the phenomenal level organize players' paths of understanding, action assumptions, and experiential structures, thereby shaping how player agency emerges. In such contexts, agency is exercised not only through acting within the system, but through players' capacity to form and revise explanations of how the system operates—what this paper refers to as explanatory agency. In this sense, AI representations in games, while not equivalent to real AI, can nevertheless provide an analogous black-box interaction structure at the phenomenal level, enabling players to develop orientations toward understanding AI through trial and error and interpretation.

From a semiotic perspective, this gap should not be seen as a flaw, but rather as a precondition for meaning-making. What players encounter is never the "game system itself," but a system-as-represented constructed through interfaces, feedback, and narrative. Understanding does not arise from mastery of the complete rule set, but is generated through the formulation of hypotheses and their continual revision in action when facing incomplete or even misleading signs. This abductive interpretive process constitutes a key mechanism through which explanatory agency is enacted in opaque systems, providing a semiotic foundation for analyzing how player agency is produced under conditions of uncertainty and limited access to underlying system logic.

### *XAI and Game*

As the field of Explainable Artificial Intelligence (XAI) has developed rapidly and achieved significant technical advances, a number of representative XAI algorithms have emerged, such as LIME (Ribeiro et al., 2016), LRP ( Binder et al., 2016), DeepLIFT (Shrikumar et al., 2017). Their core objective is to compute and present the contribution of features to model outputs, either from within the model or from prediction results, in order to enhance the explainability and auditability of model decisions. However, XAI research has been considered to progress slowly in providing explanations that are genuinely useful to users (Chromik & Butz, 2021). Studies have shown that a considerable proportion of users are unable to apply XAI explanations even in relatively simple AI tasks (Narayanan et al., 2018). Nguyen and Zhu (2022) point out that many XAI systems still fail to deliver explanations that users actually find helpful, largely because these explanatory needs are often inferred subjectively by researchers from a technical perspective, rather than being directly derived from users themselves. Against this background, some scholars have proposed shifting the research focus toward explainable interfaces (EI), viewing them as a key object of study that connects technical explanations with human understanding (T. Nguyen et al., 2024).

Computer games have been explicitly identified as one potential form of explainable interface. Through game mechanisms such as exploration, comparison, and immediate feedback, players are able to gradually understand complex AI behaviors during interaction, forming an explanatory pathway that differs from traditional XAI visualizations (T. Nguyen et al., 2024) . Existing studies have attempted to use simulation games as a form of AI explanation. Villareale et al. (2024) , by mapping XAI questions (Liao et al., 2020) onto simulation game design, analyze how existing game mechanics and interfaces in simulation games convey complex model information, and propose design considerations for improving XAI explanation interfaces through game elements.

Building on this line of work, this paper further introduces XAI questions into the context of strategy games, which differ significantly from simulation games in terms of game mechanics and interface design. From this alternative perspective, it explores the potential of strategy games to provide the information required for XAI, thereby complementing and extending existing game-design-based XAI research. Accordingly, rather than treating XAI as a question of "whether the system provides explanations," this paper focuses on whether explanations are transformed into users' explanatory agency: that is, whether users are able to form understanding based on the cues provided by interfaces, and to develop new judgments and actions on this basis.

## METHODOLOGY

### *Case Game: Arknights*

When discussing player agency, it is commonly assumed that agency is primarily addressed in games that possess a sense of worldness. Only when we are embodiedly present within a game world, and are represented within it in some form, does agency tend to be discussed more frequently. Understanding the game world constructed by *Arknights* is therefore crucial. *Arknights* is a free-to-play tactical RPG & tower defense mobile game developed by the Chinese developer *Hypergryph*. The game was released between 2019 and 2020 in mainland China and other regions. The game is set in the dystopian, apocalyptic future setting of the planet Terra. Players take on the role of the masked and amnesiac "Doctor" (whose specific nickname can be freely chosen by the player and is consistently referred to as "Doctor" in this paper), who commands an "operation squad" of the organization "Rhodes Island." Rhodes Island is an organization that integrates pharmaceutical production, medical services, and defense operations. While searching for a cure for the apocalyptic disease known as "*Oripathy*," Rhodes Island must also resist invasions from other hostile organizations ("*Arknights*," 2025).

PRTS is the shipborne AI and combat assistance system used by Rhodes Island, with its full name being the *Preliminary Rhodesisland Terminal System*. In the game, PRTS functions include assisting the commander in remote battlefield command and conducting pre-battle simulations for both the commander and combat operators. It is also capable of using AI-based analysis to support infrastructure planning and operator deployment within Rhodes Island, as well as monitoring and safeguarding the physical and psychological well-being of combat and logistics personnel, ensuring that Rhodes Island can maintain efficient combat capabilities when facing unexpected situations or crises. At the end of Chapter 15 of the game, due to the awakening of the administrator "Priestess," who possesses a higher level of administrative authority than the Doctor, PRTS begins to exhibit disorder and anomalies, and after triggering a series of disruptions, ultimately becomes unbound from Rhodes Island.

In this paper, PRTS is treated as a diegetic AI representation. It is not an artificial intelligence system that can be verified at a technical level, but rather a phenomenal black box jointly constructed through interface, feedback, and narrative frameworks. Although, at the level of computational implementation, PRTS may rely on scripted mechanisms or Wizard-of-Oz-style design approaches, the analysis in this paper does not target its underlying algorithms. Instead, it focuses on the level of the system-as-represented, namely the interface structures, information configurations, and interpretive cues that players can actually experience during gameplay.

*Method*
This paper adopts a qualitative approach grounded in close reading, interface analysis, and sustained gameplay engagement with Arknights. The focus is not on algorithmic transparency at the level of code or system implementation, but on what may be called the phenomenological black box: how opacity, interpretation, and explanation are configured at the level of player experience through narrative framing, UI design, and interaction structure.

The analysis is based on extended playthroughs of the main storyline, with particular attention to missions across early chapters and the intensified confrontation in Chapter 15. Gameplay was documented through systematic screenshot capture, observation of UI feedback, and written play notes during and after play sessions. Rather than aiming to generalize to all players, the study analyzes the position and expectations constructed for an implied player , that is, the kind of player experience the system itself anticipates, shapes, and addresses (Aarseth, 2014). Analytical categories were developed through iterative thematic grouping of recurring design patterns, including (1) how the system positions the player in relation to knowledge and control, (2) how action is mediated through the interface, and (3) how explanation, uncertainty, and trust are negotiated over time. Through this process, the study examines how explanatory agency emerges not as a given affordance, but as something progressively configured, challenged, and worked through in play.

## RESULTS
Through close gameplay and interface analysis of *Arknights*, with particular attention to the sustained observation of its AI representation system PRTS, this paper illustrates how the system is simultaneously embedded in and connects the game's virtual narrative layer and the player's operational layer. As Aarseth (2011) points out, game experience itself is a composite structure: it is constituted through the interweaving of virtual and simulated elements, and under specific conditions extends toward the real level. This framework reveals the position in which players are placed within the system, and how their actions are mediated through PRTS. Building on this foundation, the analysis further focuses on how interpretive gaps are gradually amplified at both the mechanical and narrative levels, and how they reorganize players' understanding and judgment of the system.

*From "Remote Acting Subject" to "Actionable but Unverifiable"*
Mobile games characterized by expansive world-building and a focus on combat and progression typically construct a protagonist whose personality and gender traits are minimized and whose knowledge of the world is limited; the Doctor in *Arknights* is no exception. At the beginning of the game, the Doctor awakens in a state of amnesia and gradually develops a basic understanding of the world of Terra under the guidance of the companion Amiya. This setup is not merely intended to facilitate narrative immersion; rather, it synchronizes the Doctor's knowledge state with that of the player, preventing a sense of dissonance in which players are forced to inhabit a character who already possesses extensive knowledge and established relationships with other characters. In the absence of narrative tricks or retrospective revelations, the Doctor does not function as a character who possesses key information a priori, but instead serves as a cognitive projection of the player within the game world.

Under this premise, PRTS is introduced as an indispensable interface between the Doctor and the Rhodes Island system. As a core system inherited from a previous civilization, the functions of PRTS permeate multiple levels of the Rhodes Island ship, ranging from combat command and intelligence analysis to logistics and infrastructure deployment. Nearly all critical actions must be carried out through PRTS. Regardless of how players interpret PRTS at the narrative level—whether as a trustworthy assistant or as an overly pervasive and difficult-to-control presence—the Doctor can only access the management system of Rhodes Island through a "neural connection" with PRTS (see Figure 1). In this process, PRTS becomes the primary entry point through which players understand the world, acquire information, and take action.

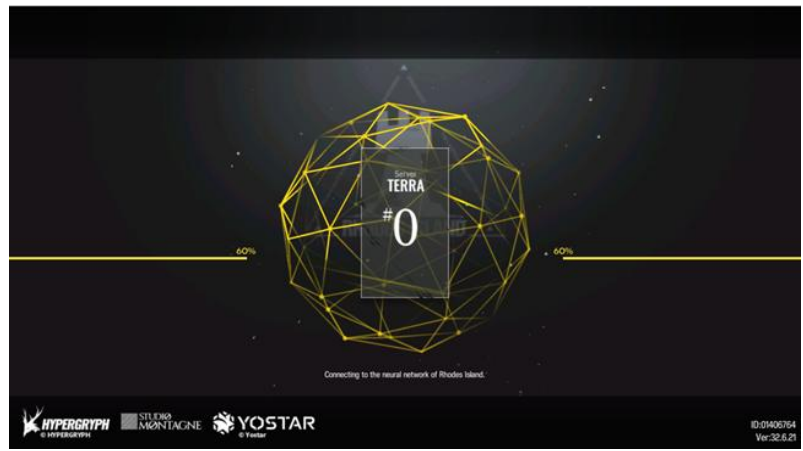

Figure 1 A login screenshot from *Arknights*, where the act of logging in is framed as a "neural connection." This interface establishes PRTS as the primary entry point through which the Doctor accesses the world, information, and action.

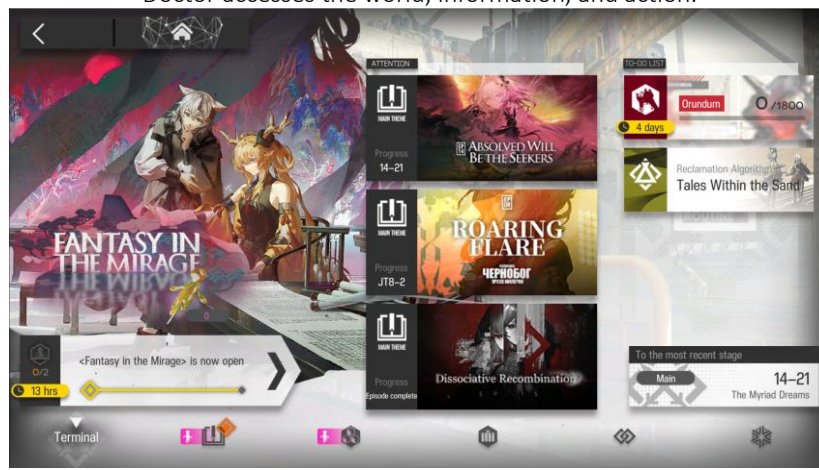

Figure 2 The terminal interface screenshot in *Arknights*, presenting multiple game modes and operational entry points. This interface visualizes how the Doctor's agency is exercised through remote command and coordination via PRTS rather than direct, embodied presence in the game world.

This mediating relationship is further reinforced at the operational level. From this perspective, the game system imposes formal constraints on player agency. As *Arknights* is not an open-world game, and as a level-based mobile game that requires continuous updates, the Doctor cannot freely traverse different locations across Terra. At the narrative level, the Doctor's physical body is situated at a specific location within the world of Terra and advances along a linear timeline of the main storyline; at the operational level, however, the Doctor repeatedly enters a series of "simulated combat" scenarios through PRTS that are not continuous with the main narrative. As a result, across multiple game modes, the Doctor does not directly intervene in the battlefield or the world in an embodied form, but instead consistently conducts remote command and coordination through the terminal (Figure 2).

Notably, following the *ver.2.5.04* update (*Terminal*, n.d.), the game's terminal interface underwent significant adjustments. On the one hand, after years of operation and the continuous introduction of new gameplay modes, the original interface structure was no longer able to accommodate the expanding content system, necessitating a UI reconstruction to reorganize level selection and operational flows. On the other hand, this interface update coincided temporally with the progression of the narrative to the end of Chapter 15. In this chapter, PRTS has already completed its "betrayal" at the narrative level, with most functional modules of Rhodes Island becoming unbound from it, and the terminal used by players being taken over by a new proxy system, ZOOT. The juxtaposition of functional changes and narrative developments leads the terminal interface to be reinterpreted at the experiential level as a form of operational mediation that is no longer neutral. Taken together, within this structure, players' "presence" is not spatial in nature, but is instead realized through flows of information, commands, and feedback.

Players' actual modes of action are likewise mediated by PRTS. At the same time, PRTS begins to intervene more deeply in the players' action process, gradually reshaping their action experience. In combat, *Arknights* primarily adopts a grid-based tower defense gameplay structure: relatively limited grid-based maps, resource systems based on automatic cost recovery and operator-generated cost, and wave-based enemy attack structures. Accordingly, it is typically classified as a strategy game with puzzle-like characteristics, in which players resist successive waves of enemies through deployment and management on the map. As a result, layout choices and the sequencing of placements become the core factors determining combat outcomes (Figure 3).

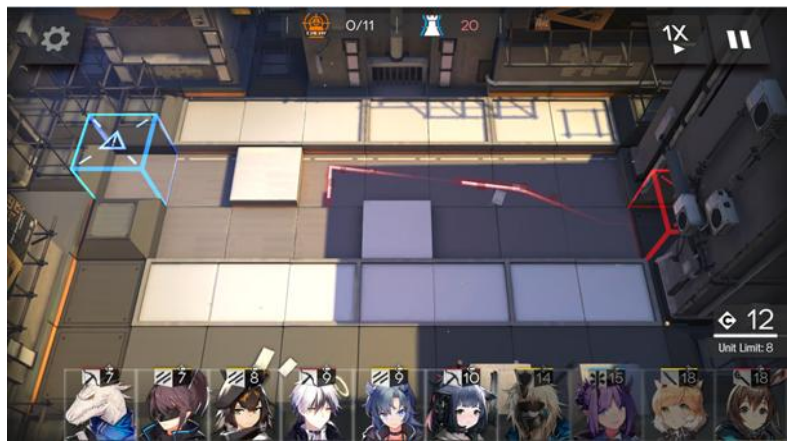

Figure 3 A combat screenshot in *Arknights,* showing the grid-based battlefield and enemy route visualization. The interface highlights how combat outcomes are structured by spatial layout and placement order rather than direct, continuous control.

Before combat begins, players first encounter an intelligence preview organized by PRTS, including enemy types, map structures, and possible movement routes; once combat starts, the system continuously updates battlefield conditions through interface feedback such as path predictions, remaining enemy counts, and remaining allied life points, providing a basis for subsequent player judgments.

Because an entire battle typically involves the parallel deployment and coordination of multiple operators, players' operational inputs do not strictly correspond to the immediate behavior of any single operator. Instead, they are primarily expressed through directive actions such as deployment, withdrawal, and skill activation. These actions are performed via dragging and clicking, and their semantics are closer to "coordination" than to "direct control." Consequently, players do not act upon battlefield units in a "soul-possession" manner, but rather translate their judgments into commands executed by the system through the interface. Their position is thus closer to that of a "behind-the-scenes orchestrator" who allocates resources and issues decisions via a computational system, rather than an agent who is personally present to carry out actions on the field.

### *"Interpretive Gaps"*

After players' actions are translated into directive decisions mediated by the system, PRTS does not provide players with a stable or reliable cognitive framework. More precisely, it offers a form of "usable explanatory representation": sufficient to support the initiation of action, yet insufficient to support the attributability and verifiability of action. As a result, players repeatedly encounter incomplete, delayed, and internally contradictory information at multiple critical moments. This instability is not an occasional malfunction, but a structural experience that runs throughout both combat procedures and narrative progression, gradually transforming "understanding the system" from a prerequisite into an ongoing labor.

First, in the pre-combat phase, PRTS typically presents enemy intelligence and route predictions as the primary basis for players' strategic planning. However, the explanatory power of such previews is always bounded. Enemy mechanics may change midway through combat; new enemies or routes may not appear in the initial intelligence; and as the narrative progresses, certain stage effects may directly interfere with the presentation of intelligence itself. Players therefore repeatedly experience moments of "surprise" in which their expectations are disrupted—not due to operational errors, but because the explanations previously provided by the system are insufficient to cover its actual operational state. During combat, players may still compensate for the lack of pre-combat information through real-time action and observation; however, once action is fully delegated to system replay, such interpretive gaps can no longer be corrected through immediate adjustment.

These interpretive gaps are further amplified in the Auto-Deploy mechanism. Auto-Deploy allows PRTS to replay previously cleared stages based on the Doctor's prior actions, enabling repeated stage completion for the purpose of farming in-game resources. However, some stages involve elements of randomness that may lead to Auto-Deploy failure. When such failures occur, the system typically presents only a result-oriented conclusion ("Auto-Deploy failed"), without indicating the specific moment or cause of failure. Players can only infer the source of deviation by replaying the stage manually or reviewing the combat process. In the absence of system-provided explanations, players must rely on repeated playthroughs and comparisons to hypothesize the source of error.

As the narrative progresses, this recurring interpretive instability at the mechanical level begins to be echoed and intensified at the narrative level. PRTS is no longer presented merely as an interface with limited information, but as an entity that may mislead, malfunction, or even undergo a shift in allegiance. Within the story, the Doctor is warned that PRTS may be suspected of rebellion, and clear fissures begin to emerge between PRTS's prompts and the judgments of other characters. Players are thus placed in a new interpretive situation: they must not only understand stage mechanics, but also judge "whether the explanations themselves are trustworthy."

The tutorial stage in Chapter 15 (15–19) can be seen as the peak of the contradiction between system and characters. Pre-combat enemy intelligence is unavailable, and the scene is explicitly marked as "information partially disrupted," yet the system still retains route prediction as a "usable clue," forcing players to oscillate between reliance and suspicion. After entering the deployment phase, PRTS repeatedly offers suggestions that indicate "reasonable positions but unreasonable orientations," and characters immediately point out through dialogue that "commands may be interfered with" and that "PRTS instructions need to be discerned," explicitly turning what could otherwise be tacitly accepted interface cues into objects requiring judgment. When players deploy Radian, PRTS provides a seemingly reasonable but ultimately failure-inducing key position: if the suggestion is followed, enemies will detour from an alternative spawn point and break through. In this moment, the only "correct action" is precisely to reject the system's recommendation and deploy the operator at the allied base entrance instead. Subsequent prompts from Rosmontis and Pith further present anomalies such as "apparently corrective yet actually ineffective" guidance and even "out-of-bounds positions that cannot be deployed," until PRTS announces that it will "disable all operator permissions" and "initiate final purge protocol," forcibly transitioning the game into Auto-Deploy. The interface then displays error messages and garbled text, allowing players to experience at the most immediate level that the system not only fails to explain, but can even revoke agency and force entry into "Auto-Deploy" (Figure 4). Here, interpretive gaps no longer manifest merely as insufficient information, but escalate into an antagonistic form: the explanations provided by the system no longer help players understand the world, but instead become interventions that must be resisted.

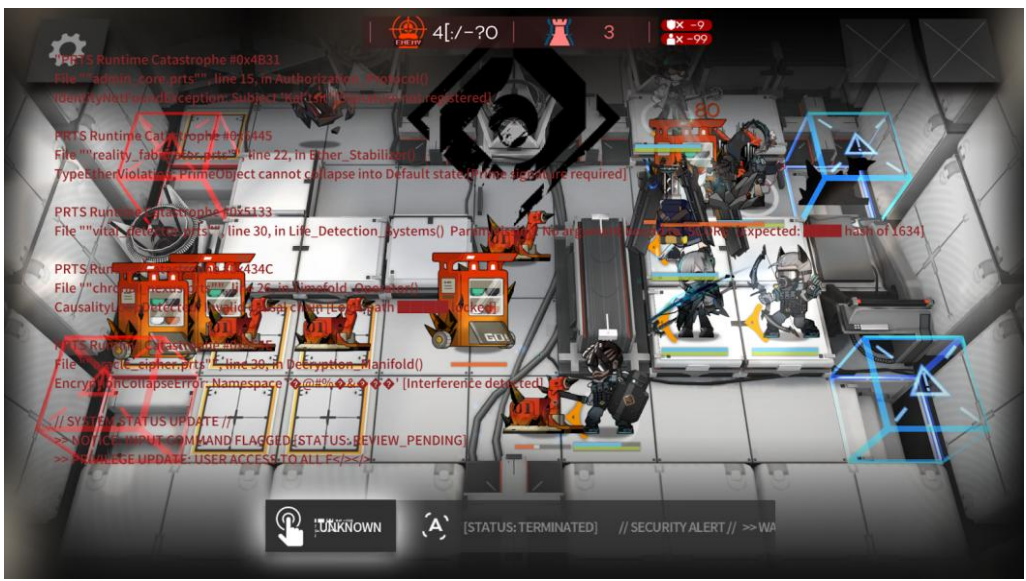

Figure 4 The 15-19 combat screenshot in *Arknights*, showing PRTS entering a failure state with error messages and forced Auto-Deploy. The interface demonstrates how system guidance shifts from incomplete explanation to direct intervention, temporarily revoking player agency.

## DISCUSSION

In *Arknights*, PRTS does not exist merely as a functional interface. Through narrative construction, mediation of core gameplay, and the strategic withholding of explanation, it gradually configures a composite form of implied player agency. The results presented above show that player agency is not simply manifested in "whether one can take action" or "whether one has freedom of choice," but is systematically organized in the positioning of action within the game world, the ways in which action affects that world, and the responsibility of interpretation under conditions of system opacity.

Through an analysis of the PRTS system in *Arknights*, we can observe how a form of agency centered on interpretation is constructed within players' gameplay experience.

### *Embodiment and Command under Mediated Relations*

For players, PRTS functions like an interface bridge, translating aspects of the world of Terra that the Doctor cannot directly experience into "readable texts," and enabling otherwise static game elements—such as Operators, ship spaces, and infrastructure facilities—to respond to input and generate feedback. Operators' abilities are translated into numerical indicators such as experience points and attack values, while ship rooms and furniture are quantified as parameters of efficiency, value, and comfort. As a result, the game world is not experienced in a first-hand, embodied manner, but is presented through a systematized, filtered, and recoded form of information. In other words, PRTS reconfigures the meaning-configuration of the game world and rebalances the boundaries of player agency.

From the perspective of embodiment relation, PRTS's role in combat is not limited to serving as a mediating interface between the player and the battlefield; rather, it becomes further embedded in the player's action structure, forming part of the player's "acting body." Players do not first formulate an independent and complete intention and then execute it through the interface. Instead, their judgments, expectations, and actionable possibilities are already pre-shaped by the operational logic and feedback mechanisms configured by PRTS from the very beginning of combat. As with embodied technologies more generally, PRTS does not appear as an explicitly perceived object, but enters action through its "disappearing" quality, allowing players to act through technology rather than toward technology at a pre-reflective level. In this sense, player agency is not diminished by PRTS, but reorganized by it, and presented as a form of directive bodily intentionality premised on the system. Players are still able to act, yet the manner, rhythm, and consequences of their actions no longer remain fully within the scope of direct perception or verification, and must instead be indirectly confirmed through system operation and feedback. This state of being "actionable but unverifiable" reveals how technology reshapes action experience within embodiment relations: action is no longer equivalent to direct intervention in the world, but is delayed, distributed, and ultimately validated through a system with autonomous execution capacity.

PRTS thus not only fulfills functional roles at the operational level, but also reorganizes the player–world relationship at the experiential level. The world is no longer an object directly perceived, but one that must be understood and acted upon through system mediation. Accordingly, players are presupposed as subjects who must make judgments and decisions within the system, where the precondition of agency lies not in complete control over the system, but in sustained reliance on the cognitive channels it provides.

### *Explanatory Agency and Trust Calibration under Opaque Systems*

PRTS does not provide a stable or reliable cognitive framework, but rather a form of "usable explanatory representation": sufficient to initiate action, yet insufficient to support the attributability and verifiability of action. Across combat previews, Auto-Deploy, and later stages, players repeatedly encounter incomplete, delayed, and internally contradictory information. The key issue here is not information incompleteness per se, but how it alters players' attribution of feedback. Failure can no longer be easily reduced to individual judgment errors, but is instead suspended within the question of "whether the system has provided sufficient information." In this process, PRTS does not assume responsibility for explaining "why failure occurred," but instead outsources interpretive labor to the player, who must generate working hypotheses about the system through trial, error, and comparison. Players are still able to act, yet struggle to stably connect action and outcome into verifiable causal chains. Player agency is thus distilled into a form of explanatory agency: the capacity to maintain effective decision-making and actively calibrate one's cognitive model of the system under conditions of opacity.

Gameplay and narrative do not simply form ludonarrative harmony or ludonarrative dissonance, but instead constitute a structure of tension. At the mechanical level, players must continue to rely on PRTS prompts to maintain operational fluency, while the narrative repeatedly warns that such reliance may lead to error and loss of control. Within this structure, players are compelled into a metacognitive state: they must not only understand level logic, but also judge "whether the explanation itself is trustworthy." Understanding thus ceases to be a prerequisite for action, and instead becomes a continuously generated and repeatedly tested outcome of action.

When situated within the context of XAI research, these findings suggest that *Arknights* offers a perspective distinct from traditional explanatory paradigms. Conventional XAI research often seeks to eliminate uncertainty by increasing algorithmic transparency, yet this can lead to users' cognitive passivity when faced with incorrect explanations. The case of PRTS demonstrates that explanation does not necessarily need to be presented as a static "answer," but can instead function as a playable process. Through strategic withholding, delayed feedback, and narrative disruptions of trust, systems can guide users to shift from passive recipients of information to active practitioners of interpretation. This paradigm shift from "result-oriented" to "process-oriented" explanation not only deepens users' understanding of complex systems, but also fosters a form of reflective agency within human–machine collaboration.

## CONCLUSION

In conclusion, through an analysis of the PRTS system in Arknights, this paper proposes a pathway for understanding player agency centered on explanatory agency. The contribution of this study does not lie in providing new empirical examples for existing theories, but in revealing a phenomenon that has been less systematically examined: how interpretive gaps can be designed as playable structures, thereby reshaping players' agentic experience. This analysis extends the concept of agency within game studies, offers design implications for XAI and explainable interface research that foreground interpretive processes rather than explanatory outcomes, and demonstrates the distinctive methodological value of digital games for exploring how people understand complex systems.

## ACKNOWLEDGEMENTS
This work originated from a research-oriented course project conducted during my Erasmus+ exchange at the IT University of Copenhagen and is part of my independent academic research as a master's student at Uppsala University.

## AI USAGE STATEMENT
During the writing of this paper, ChatGPT 5.2 was used to assist with the translation of certain professional terms and to refine academic style and logical clarity in parts of the manuscript. All conceptual contributions are solely the author's.